\documentclass{article}
\usepackage{amssymb}
\usepackage{bm}
\usepackage[dvips]{graphicx}
\begin{document}
\title{Disturbance of spin equilibrium by current\\
through the interface of noncollinear ferromagnets}
\author{E.~M.~Epshtein, Yu.~V.~Gulyaev, P.~E.~Zilberman\thanks{E-mail: zil@ms.ire.rssi.ru}
\\ \\ Institute of Radio Engineering and Electronics \\
Russian Academy of Sciences \\ Fryazino, Moscow Region, 141190, Russia}
\date{}
\maketitle
\begin{abstract}
Boundary conditions are derived that determine the penetration of spin
current through an interface of two non-collinear ferromagnets with an
arbitrary angle between their magnetization vectors. We start from the
well-known transformation properties of an electron spin wave functions
under the rotation of a quantization axis. It allows directly find the
connection between partial electric current densities for different spin
subbands of the ferromagnets. No spin scattering is assumed in the near
interface region, so that spin conservation takes place when electron
intersects the boundary. The continuity conditions are found for partial
chemical potential differences in the situation. Spatial distribution of
nonequilibrium electron magnetizations is calculated under the spin
current flowing through a contact of two semi-infinite ferromagnets. The
distribution describes the spin accumulation effect by current and
corresponding shift of the potential drop at the interface. These effects
appear strongly dependent on the relation between spin contact resistances
at the interface.
\end{abstract}

\section{Introduction}\label{section1}
A branch of the solid state physics and electronics called
``spintronics''~\cite{Heinrich} developed rapidly last years. The name is
due to the decisive role which the electron spin and related magnetic
moment play in the transport phenomena studied. The most important
phenomena appear in magnetic junctions containing ferromagnetic layers.
Disturbance of spin equilibrium occurs when spin-polarized current flows
through the interfaces of the layers. The disturbance leads to a number of
new spin dependent contact phenomena, which are of interest for a theory
development and for using under interpretation of experimental data.

To solve equations of motion for the junctions, we should have true
boundary conditions at the interfaces. The problem of boundary conditions
appears here by a natural way. The most of the published works consider
either a contact between ferromagnetic and nonmagnetic
materials~\cite{vanSon}--\cite{Lee} or between two ferromagnets with
collinear magnetic moments such as a domain wall~\cite{Zvezdin}. In the
both cases, there is a single quantization axis. Meanwhile, the boundary
conditions determining spin current through a contact of two noncollinear
ferromagnets are significant for a number of problems concerning
spin-polarized current induced spin switching~\cite{Gulyaev1}. Such type
of boundary conditions are treated in the present work.

Using the boundary conditions derived, we calculated further the following
contact phenomena: magnetization distribution, spin accumulation shift of
the contact potential drop and spin accumulation contribution to contact
resistance.

\section{Boundary conditions for magnetization flux}\label{section2}
The electron magnetization distribution $m_i(x,\,t)\;(i=x,\,y,\,z)$ is
described by the continuity equation
\begin{equation}\label{1}
  \frac{\partial m_i}{\partial t}+\nabla_kJ_{ik}=-\frac{m_i-\bar
  m_i}{\tau},
\end{equation}
where $\bar m_i$ is equilibrium value of the electron magnetization,
$\tau$ is spin relaxation time, $J_{ik}$ is the electron magnetization
flux density (the first index determines the magnetization vector
direction in the flux, the second one indicates the flux propagation
direction).

Basing on the well known derivation of the quantum mechanical formula for
the particle (electron) current density~\cite{Landau}, we obtain the
electron magnetization flux density in the following form:
\begin{eqnarray}\label{2}
  &&J_{ik}\bigl(\mathbf r,\,t\bigr)=\frac{i\hbar\mu_B}{2m}\sum_{\mathbf
  p,\,s_1,\,s_2}\sigma_i^{s_1s_2}\Bigl(\psi_{\mathbf p,\,s_2}\bigl(\mathbf
  r,\,t\bigr)\nabla_k\psi_{\mathbf p,\,s_1}^\ast\bigl(\mathbf
  r,\,t\bigr)\nonumber \\
  &&-\psi_{\mathbf p,\,s_1}^\ast\bigl(\mathbf
  r,\,t\bigr)\nabla_k\psi_{\mathbf p,\,s_2}\bigl(\mathbf
  r,\,t\bigr)\Bigr),
\end{eqnarray}
where $\psi_{\mathbf p,\,s}(\mathbf r,\,t)$ is electron wavefunction with
momentum $\mathbf p$ and spin state $s$, $\mu_B$ is Bohr magneton,
$\boldsymbol\sigma=\{\sigma_x,\,\sigma_y,\,\sigma_z\}$ are Pauli matrices.

Let us see how the magnetization flux transforms under rotation of the
quantization axis. Such a rotation can be due to electron transfer from
one to another magnetic layer of the junction with different direction of
the quantization axis as well as rotation of a homogeneous medium
quantization axis (by an external magnetic field, for example).

Let electric current flow along $x$ axis with quantization axis parallel
to $z$ axis. The magnetization flux density has single component $J_{zx}$.
Then the electron current goes into another layer with quantization axis
parallel to $z'$ axis that makes an angle $\chi$ with $z$ axis. The
quantization axis rotation is described by a spin wave function
transformation matrix~\cite{Landau}
\begin{equation}\label{3}
\hat U_x(\chi)=\left(\matrix{
  \cos(\chi /2) & i\sin(\chi /2) \cr
    i\sin(\chi /2) & \cos(\chi /2)}\right).
\end{equation}

Such a transformation of the wave functions leads to transformation of the
magnetization flux density, so that a longitudinal component
$J_{z'x}=J_{zx}\cos\chi$ appears with polarization along the new
quantization axis $z'$ as well as a transverse component
$J_{y'x}=J_{zx}\sin\chi$ with perpendicular polarization. Very different
spin relaxation times correspond to the longitudinal and transverse
polarizations, so only the longitudinal component $J_{z'x}=J_{zx}\cos\chi$
survives beyond a thin layer of Fermi wavelength thickness (the so called
Berger--Slonczewski layer, see~\cite{Gulyaev1} for details). This gives a
boundary condition for the electron magnetization flux density at the
interface between the ferromagnets $x=0$:
\begin{equation}\label{4}
  J_{zx}\cos\chi\Bigr|_{x=-0}=J_{z'x}\Bigr|_{x=+0}.
\end{equation}

Let us see how the partial spin-polarized current densities for spin-up
and spin-down electrons transform under changing the quantization axis. We
have
\begin{equation}\label{5}
  j_++j_-=j,
\end{equation}
\begin{equation}\label{6}
  j_+-j_-=\frac{e}{\mu_b}J_{zx},
\end{equation}
where $j$ is total current density. From Eqs.~(\ref{5}) and~(\ref{6}) we
obtain
\begin{equation}\label{7}
  j_\pm=\frac{1}{2}\left(j\pm\frac{e}{\mu_B}J_{zx}\right).
\end{equation}

The current density $j$ does not change under the quantization axis
rotation, while the magnetization flux density $J_{zx}$ transforms in
accordance with Eq.~(\ref{4}). Therefore, the transformed partial current
densities take the form
\begin{equation}\label{8}
  j'_\pm=\frac{1}{2}\left(j\pm\frac{e}{\mu_B}J_{zx}\cos\chi\right).
\end{equation}
With Eqs.~(\ref{5}) and~(\ref{6}) taking into account, a transformation
law for the partial current densities takes the form
\begin{equation}\label{9}
  j'_\pm=j_\pm\cos^2\frac{\chi}{2}+j_\mp\sin^2\frac{\chi}{2}.
\end{equation}

The electric current transformations~(\ref{9}) were obtained previously
in~\cite{Berger}--\cite{Gulyaev3} by other ways.

\section{Boundary conditions for chemical potentials}\label{section3}
The other boundary condition is imposed on the partial chemical potentials
$\zeta_\pm$ corresponding to spin-up and spin-down electrons. In its
derivation, we start from the energy flux continuity condition at the
interface between two ferromagnets.

The Boltzmann equation for spin-up and spin-down electron distribution
functions $f_{\mathbf p\pm}$ takes the form
\begin{equation}\label{10}
  \frac{\partial f_{\mathbf p\pm}}{\partial t}
  +\mathbf v_\pm(\mathbf p)\frac{\partial f_{\mathbf p\pm}}{\partial\mathbf r}
  -e\frac{\partial\varphi(\mathbf r)}{\partial\mathbf r}
  \frac{\partial f_{\mathbf p\pm}}{\partial\mathbf p}=I_{\mathbf p\pm},
\end{equation}
where $\varphi(\mathbf r)$ is electric potential, $\mathbf v_\pm(\mathbf
p)=\partial\varepsilon_{\mathbf p\pm}/\partial\mathbf p$ is electron
velocity, $\varepsilon_{\mathbf p\pm}$ is the electron energy with
momentum $\mathbf p$, $I_{\mathbf p\pm}$ is collision integral.

Multiplying Eq.~(\ref{10}) by electron energy $\varepsilon_{\mathbf p\pm}$
and summing over $\mathbf p$ gives energy conservation law
\begin{equation}\label{11}
  \frac{\partial U_\pm}{\partial t}+\mathrm{div}\mathbf W_\pm
  =\sum_{\mathbf p}\varepsilon_{\mathbf p\pm}I_{\mathbf p\pm},
\end{equation}
where $U_\pm=\sum\limits_{\mathbf p}\varepsilon_{\mathbf p\pm}f_{\mathbf
p\pm}$ are partial electron energy densities,
\begin{equation}\label{12}
  \mathbf W_\pm=\sum_{\mathbf p}\left(\varepsilon_{\mathbf p\pm}
  +e\varphi\right)\mathbf v_\pm\left(\mathbf p\right)f_{\mathbf p\pm}
  =\frac{1}{e}\mathbf j_\pm\left(\zeta_\pm+e\varphi\right)
\end{equation}
are partial energy flux densities for completely degenerate electrons.

Consider a contact in $x=0$ plane between two homogeneous semi-infinite
ferromagnets with different quantization axes. The total energy flux
density along $x$ axis is
\begin{eqnarray}\label{13}
  W=W_++W_-=\frac{1}{e}\left[j_+\left(\zeta_++e\varphi\right)
  +j_-\left(\zeta_-+e\varphi\right)\right]\nonumber \\
=\frac{1}{2e}j\left(\zeta_++\zeta_-+2e\varphi\right)
+\frac{1}{2\mu_B}J_{zx}\left(\zeta_+-\zeta_-\right),
\end{eqnarray}
where $J_{zx}$ is the electron magnetization flux density; we used
Eq.~(\ref{7}) here.

With boundary condition~(\ref{4}) taking into account, the energy flux
continuity condition $W\bigr|_{x=-0}= W'\bigr|_{x=+0}$ takes the form
\begin{eqnarray}\label{14}
    j\left\{\left(\zeta_++\zeta_-+2e\varphi\right)\Bigr|_{x=-0}-\left(\zeta'_+
    +\zeta'_-+2e\varphi\right)\Bigr|_{x=+0}\right\}\nonumber \\
+\frac{e}{\mu_B}J_{zx}\left\{\left(\zeta_+-\zeta_-\right)\Bigr|_{x=-0}-\left(\zeta'_+
-\zeta'_-\right)\Bigr|_{x=+0}\cos\chi\right\}=0.
\end{eqnarray}

The form of the boundary conditions should not depend on values of the
current and magnetization flux. Therefore, the contents of both curly
brackets in Eq.~(\ref{14}) are to vanish each separately. This gives the
following boundary condition for the partial chemical potentials at the
interface:
\begin{equation}\label{15}
\left(\zeta_+-\zeta_-\right)\Bigl|_{x=-0}=\left(\zeta'_+-\zeta'_-\right)\Bigr|_{x=+0}\cos\chi.
\end{equation}

From Eq.~(\ref{15}), a boundary condition can be obtained for
nonequilibrium electron magnetization $\Delta m=\mu_B\left(\Delta
n_+-\Delta n_-\right)$, where $\Delta n_\pm$ are deviations of the partial
electron densities in spin subbands from their equilibrium values $\bar
n_\pm$. Because of the neutrality condition we have $\Delta n_++\Delta
n_-=0$, so that $\Delta n_\pm=\pm\Delta m/2\mu_B$. The partial chemical
potentials are related with $\Delta n_\pm$, namely,
\begin{equation}\label{16}
  \zeta_\pm-\bar\zeta=\frac{\Delta n_\pm}{g_\pm}
  =\pm\frac{\Delta m}{2\mu_Bg_\pm},\quad\zeta'_\pm-\bar\zeta'
  =\frac{\Delta n'_\pm}{g'_\pm}=\pm\frac{\Delta m'}{2\mu_Bg'_\pm},
\end{equation}
where $g_\pm,\;g'_\pm$ are the densities of states at the Fermi level for
spin-up and spin-down electrons, $\bar\zeta,\;\bar\zeta'$ are the
equilibrium chemical potentials of two ferromagnets.

With Eq.~(\ref{16}) taking into account, we find
\begin{equation}\label{17}
  \zeta_+-\zeta_-=N\Delta m,\quad N=\frac{1}{2\mu_B}\left(\frac{1}{g_+}+\frac{1}{g_-}\right).
\end{equation}

From Eqs.~(\ref{15}) and~(\ref{17}), we obtain the following boundary
condition for nonequilibrium electron magnetization:
\begin{equation}\label{18}
  N\Delta m\bigl|_{x=-0}=N'\Delta m'\bigr|_{x=+0}\cos\chi.
\end{equation}

The boundary conditions~(\ref{4}),~(\ref{15}) and~(\ref{18}) correspond to
current flowing in the positive direction of $x$ axis. Under opposite
current direction, the conditions may be analogously presented in the form
\begin{equation}\label{19}
  J_{zx}\bigl|_{x=-0}=J_{z'x}\bigr|_{x=+0}\cos\chi,
\end{equation}
\begin{equation}\label{20}
  (\zeta_+-\zeta_-)\bigl|_{x=-0}\cos\chi=(\zeta'_+-\zeta'_-)\bigl|_{x=+0},
\end{equation}
\begin{equation}\label{21}
  N\Delta m\bigl|_{x=-0}\cos\chi=N'\Delta m'\bigl|_{x=+0}.
\end{equation}

\section{Electron magnetization distribution}\label{section4}
As an illustration, let us consider spin flux transfer through a contact
between two semi-infinite noncollinear ferromagnets in plane $x=0$. The
electron magnetization distribution is described by Eq.~(\ref{1}) with
magnetization flux density
\begin{equation}\label{22}
  J_{zx}=\frac{\mu_B}{e}Qj-\tilde D\frac{\partial(\Delta m)}{\partial x},
\end{equation}
where $\tilde D=(D_+\sigma_-+D_-\sigma_+)/\sigma$ is effective spin
diffusion constant, $D_\pm$ are partial electron diffusion constants,
$\sigma_\pm$ are partial conductivities, $\sigma=\sigma_++\sigma_-$ is
total conductivity, $Q=(\sigma_+-\sigma_-)/\sigma$ is conduction
polarization coefficient (see~\cite{Gulyaev1} for details). The stationary
solution of Eq.~(\ref{1}) with boundary conditions~(\ref{4})
and~(\ref{18}) takes the form
\begin{equation}\label{23}
  \Delta
  m(x<0)=\mu_Bn\frac{j}{j_D}\frac{(Q\cos\chi-Q')\cos\chi}{\nu+\cos^2\chi}\exp(x/l),
\end{equation}
\begin{equation}\label{24}
  \Delta
  m'(x>0)=\mu_Bn'\frac{j}{j'_D}\frac{(Q\cos\chi-Q')\nu}{\nu+\cos^2\chi}\exp(-x/l'),
\end{equation}
where $j_D=enl/\tau$, $l=\sqrt{\tilde D\tau}$ is spin diffusion length,
$\nu=\left(j'_D/j_D\right)\left(N/N'\right)$; the current flows along
positive direction of $x$ axis. With relationship
$\sigma_\pm=e^2D_\pm g_\pm$~\cite{Abrikosov} taking into account, the parameter $\nu$ can
be represented as
\begin{equation}\label{25}
  \nu=\frac{Z}{Z'},\quad
  Z=\frac{l}{\sigma(1-Q^2)}.
\end{equation}

Quantity $Z$ has dimensionality of contact resistance
($\rm{Ohm}\times\rm{cm}^2$), so that the parameter $\nu$ that determines
spin current matching between two ferromagnets may be treated as a ratio
of ``spin resistances''.

\begin{figure}
\includegraphics{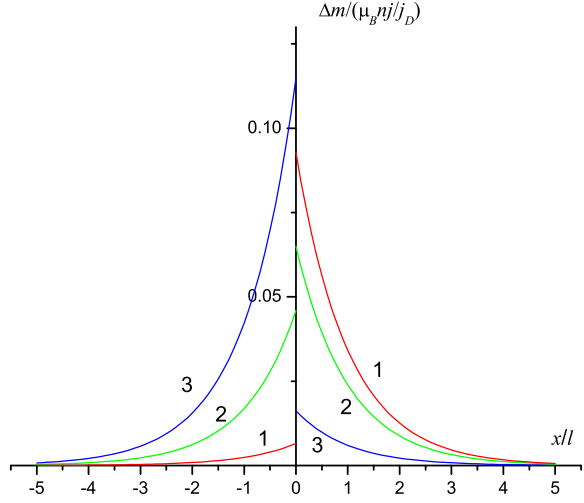}
\caption{The nonequilibrium magnetization distribution (in dimensionless
variables) at different values of the spin resistance ratio $\nu$: 1~---
$\nu=10$ (red), 2~--- $\nu=1$ (green), 3~--- $\nu=0.1$ (blue).
$Q=0.35,\;Q'=0.15,\;
l=l',\;n=n',\;j_{D}=j'_{D},\;\chi=45^\circ$.}\label{fig1}
\end{figure}

At $\nu\gg 1$, the cathode layer works as an ideal injector with
equilibrium spin polarization ($\Delta m=0$), while equilibrium breaks in
the anode layer. In the opposite case, $\nu\ll 1$, ideal collector regime
takes place, in which spin equilibrium breaks in the cathode layer. The
nonequilibrium magnetization distribution at difference values of $\nu$
parameter is shown in Fig.~\ref{fig1}.

\section{Spin accumulation resistance}\label{section5}
The spin equilibrium disturbance contributes to the resistance of the
system in study. We have
\begin{equation}\label{26}
  j_\pm=-\sigma_\pm\left(\frac{d\varphi}{dx}+\frac{1}{e}\frac{d\zeta_\pm}{dx}\right).
\end{equation}
With Eqs.~(\ref{5}) and~(\ref{16}) taking into account,
\begin{equation}\label{27}
  \frac{d\varphi}{dx}=-\frac{1}{\sigma}\left[j+\frac{1}{e}\left(\sigma_+\frac{d\zeta_+}{dx}
  +\sigma_-\frac{d\zeta_-}{dx}\right)\right]
  =-\frac{1}{\sigma}\left[j+\frac{e}{2\mu_B}\left(D_+-D_-\right)\frac{dm}{dx}\right].
\end{equation}

By integrating Eq.~(\ref{27}) over $x$ with potential drop at $x=0$ taking
into account, we obtain
\begin{eqnarray}\label{28}
  &&\varphi(-0)-\varphi(-L)+\varphi(L')-\varphi(+0)=-\frac{j}{\sigma}L
  -\frac{e}{2\mu_B\sigma}\left(D_+-D_-\right)\Delta m(-0)\nonumber \\
&&-\frac{j}{\sigma'}L'-\frac{e}{2\mu_B\sigma'}\left(D'_+-D'_-\right)\Delta
m'(+0),
\end{eqnarray}
where $L,\,L'$ are thicknesses of the contacting layers ($L\gg l,\;L'\gg
l'$).

The total potential drop over the whole system is
\begin{eqnarray}\label{29}
  &&U\equiv\varphi(-L)-\varphi(L')=\left(\frac{L}{\sigma}
  +\frac{L'}{\sigma'}\right)j+\left[\varphi(-0)-\varphi(+0)\right]\nonumber
  \\ &&+\frac{e}{2\mu_B}\left[\frac{1}{\sigma}\left(D_+-D_-\right)\Delta
  m(-0)-\frac{1}{\sigma'}\left(D'_+-D'_-\right)\Delta m'(+0)\right].
\end{eqnarray}

By equating the content of the first curly brackets in Eq.~(\ref{14}) to
zero, we find the part of the potential drop at the interface:
\begin{eqnarray}\label{30}
  &&\varphi(-0)-\varphi(+0)=\frac{1}{2e}\left(\zeta'_++\zeta'_--\zeta_+-\zeta_-\right)\nonumber \\
  &&=\frac{1}{e}(\bar\zeta'-\bar\zeta)+\frac{1}{2e}\left(\Delta\zeta'_++\Delta\zeta'_-
  -\Delta\zeta_+-\Delta\zeta_-\right)=\frac{1}{e}(\bar\zeta'-\bar\zeta)\nonumber \\
&&+\frac{e}{4\mu_B}\left[\left(\frac{D'_+}{\sigma'_+}
-\frac{D'_-}{\sigma'_-}\right)\Delta m'(+0) -\left(\frac{D_+}{\sigma_+}
-\frac{D_-}{\sigma_-}\right)\Delta m(-0)\right].
\end{eqnarray}

By substituting Eq.~(\ref{30}) into~(\ref{29}), we get after some
manipulations
\begin{equation}\label{31}
  U=U_0+\frac{e}{\mu_B}\left[\frac{Q\tilde DZ}{l}\Delta m(-0)
  -\frac{Q'\tilde D'Z'}{l'}\Delta m'(+0)\right],
\end{equation}
where
\begin{equation}\label{32}
  U_0=\left(\frac{L}{\sigma}+\frac{L'}{\sigma'}\right)j
  +\frac{1}{e}\left(\bar\zeta'-\bar\zeta\right)
\end{equation}
is the voltage in absence of the spin equilibrium breaking.

Substitution of Eqs.~(\ref{23}) and~(\ref{24}) into Eq.~(\ref{31}) gives
\begin{equation}\label{33}
  U=U_0+jZ\frac{\left(Q\cos\chi-Q'\right)^2}{\nu+\cos^2\chi}.
\end{equation}
\begin{figure}[h]
\includegraphics{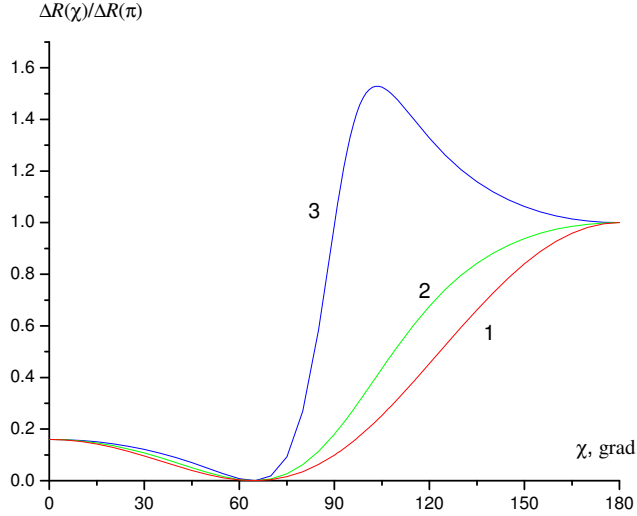}
\caption{Spin accumulation resistance as a function of the angle $\chi$
between the layer magnetization vectors at different values of the spin
resistance ratio $\nu$: 1~--- $\nu=10$ (red), 2~--- $\nu=1$ (green), 3~---
$\nu=0.1$ (blue). $Q=0.35,\;Q'=0.15$.}\label{fig2}
\end{figure}
The contribution of spin accumulation to the resistance as a function of
$\chi$ angle can be found:
\begin{equation}\label{34}
  \Delta R(\chi)\equiv\frac{U-U_0}{j}=ZZ'\frac{(Q\cos\chi-Q')^2}{Z+Z'\cos^2\chi}.
\end{equation}

The angular dependence of $\Delta R$ is shown in Fig.~\ref{fig2}. Note
that $\Delta R$ vanishes at $\chi=\arccos(Q'/Q)$.

The results obtained may be considered as a generalization of those in
Refs.~\cite{Valet,Zvezdin,Zhang} to the case of nonidentical noncollinear
ferromagnets. Really, in Ref.~\cite{Zvezdin} the contact potential drop
was calculated, that is, in essence, the first square bracket in
Eq.~(\ref{29}). On the other hand, in Ref.~\cite{Zhang} the volume
potential difference, that is the last square bracket in Eq.~(\ref{29})
was calculated. As it is seen from Eq.~(\ref{29}), we should sum the
brackets to obtain the final result. In addition, we take an arbitrary
angle $\chi$ instead of collinear orientation taken in
Refs.~\cite{Zvezdin} and~\cite{Zhang}.

Giant magnetoresistance (GMR) can be found from Eq.~(\ref{34}):
\begin{equation}\label{35}
  \mathrm{GMR}\equiv\frac{\Delta R(\pi)-\Delta R(0)}{\Delta
  R(\pi)}=\frac{4QQ'}{(Q+Q')^2}.
\end{equation}

At $Q=Q'$ we have $\Delta R(0)=0$, so that GMR takes its maximum value
$\mathrm{GMR}=1$.

\section{Conclusion}\label{section6}
We show the longitudinal spin flux continuity at the junction interface
follows directly from the spin transformation properties under the
rotation of a quantization axis.

We derive for the first time the continuity conditions of mobile electron chemical potentials at the
interface of two ferromagnetic junction layers having an arbitrary angle
between their magnetization vectors. When the conditions were derived, the
only statement was employed significantly, namely, the conservation low of
the mobile electron energy flux density at the interface.

Electron magnetization distribution in the junction is calculated based on the
boundary conditions derived. Matching parameter is discussed, which
determine the spin flux penetration through the interface. The parameter
may be treated as a ratio of the contacting layers spin resistances.

We show the disturbance of spin equilibrium at the interface leads to angle dependent shift of a
contact potential drop and to spin accumulation magnetoresistance. The
last effect was numerically estimated and the angles are found of minimal
and maximal magnetoresistance.

\section*{Acknowledgements}
The authors are thankful to Professor Sir Roger Elliott for attraction of
their attention to the problem of ferromagnetic junction switching and for
fruitful collaboration.

The work was supported by Russian Foundation for Basic Research, Grant
No.~06-02-16197.


\begin{thebibliography}{99}
\bibitem{Heinrich}
B.~Heinrich, {\em Canad. J. Phys.} {\bf 78}, 161 (2000).
\bibitem{vanSon}
P.~C.~van Son, H.~van Kempen, P.~Wyder, {\em Phys. Rev. Lett.} {\bf 58},
2271 (1987).
\bibitem{Valet}
T.~Valet, A.~Fert, {\em Phys. Rev. B} {\bf 48}, 7099 (1993).
\bibitem{Smith}
D.~L.~Smith, R.~N.~Silver, {\em Phys. Rev. B} {\bf 64}, 045323 (2001).
\bibitem{Lee}
B.~C.~Lee, {\em J. Korean Phys. Soc.} {\bf 47}, 1093 (2005).
\bibitem{Zvezdin}
A.~K.~Zvezdin, K.~A.~Zvezdin, {\em Bulletin of the Lebedev Physics
Institute} No. 8, 3 (2002).
\bibitem{Gulyaev1}
Yu.~V.~Gulyaev, P.~E.~Zilberman, E.~M.~Epshtein, R.~J.~Elliott, {\em J.
Exp. Theor. Phys.} {\bf 100}, 1005 (2005).
\bibitem{Landau}
L.~D.~Landau, E.~M.~Lifshitz, {\em Quantum Mechanics (Non-Relativistic
Theory)}, Pergamon Press, London, 1977.
\bibitem{Berger}
L.~Berger, {\em Phys. Rev. B} {\bf 54}, 9353 (1996).
\bibitem{Utsumi}
Y.~Utsumi, Y.~Shimizu, H.~Miyazaki, {\em J. Phys. Soc. Japan} {\bf 68},
3444 (1999).
\bibitem{Gulyaev2}
Yu.~V.~Gulyaev, P.~E.~Zilberman, E.~M.~Epshtein, R.~J.~Elliott, {\em J.
Commun. Technol. and Electronics} {\bf 48}, 942 (2003).
\bibitem{Gulyaev3}
E.~J.~Elliott, E.~M.~Epshtein, Yu.~V.~Gulyaev, P.~E.~Zilberman, {\em J.
Magn. Magn. Mater.} {\bf 271}, 88 (2004).
\bibitem{Gulliere}
M.~Gulliere, {\em Phys. Lett. A} {\bf 54}, 225 (1975).
\bibitem{Abrikosov}
A.~A.~Abrikosov, {\em Fundamentals of the Theory of Metals}, North-Holland
Publ. Comp., Amsterdam, 1988.
\bibitem{Zhang}
S.~Zhang, P.~M.~Levy, {\em Phys. Rev. B} {\bf 65}, 052409 (2002).
\end{thebibliography}
\end{document}